\documentclass{svproc}

\usepackage{graphicx}
\usepackage{subfigure}

\begin{document}
\title{An Experiment Study on Federated Learning Testbed}
%
%
\author{Cheng Shen
 \inst{1,2} \and Wanli Xue\inst{1,2}\thanks{Corresponding author.}}
\authorrunning{C. Shen and W. Xue}
%
\institute{The University of New South Wales, UNSW 2052 NSW, Australia \and
Cyber Security Cooperative Research Centre, Australia  \\
\email{ \{ z5141506, w.xue\} @unsw.com}\\
}
\maketitle              
\begin{abstract}
While Internet of Things (IoT) can benefit from machine learning by outsourcing model training on the cloud, user data exposure to an untrusted cloud service provider can pose threat to user privacy. Recently, federated learning is proposed as an approach for privacy-preserving machine learning (PPML) for the IoT, while its practicability remains unclear. This work presents the evaluation on efficiency and privacy performance of a readily available federated learning framework based on PySyft, a Python library for distributed deep learning. It is observed that training speed of the framework is significantly slower that of the centralized approach due to communication overhead. Meanwhile, the framework bears some vulnerability to potential man-in-the-middle attacks at network level. The report serves as a starting point for PPML performance analysis and suggests the future direction for PPML framework development.

\end{abstract}
\section{A Novel PPML Framework: Federated Learning}
\vspace{-0.05in}
As a powerful computational tool, machine learning enables devices to optimize statistical models without explicit instructions and provides various forms of automation. However, it is still challenging to apply machine learning algorithm to IoT systems given the limited computation resources and power supply on IoT devices. It is thus common for IoT systems to outsource machine-learning-based model training from the cloud,
where a centralized server acquires raw data from IoT devices to perform optimization on models.
However, this conventional method  is causing new privacy concerns. Raw user data is directly exposed to the cloud service provider, which in many cases is an untrusted third party. Due to this fact, measures have already been carried out to restrict related applications of machine learning, where potential privacy disclosure poses threat to public privacy. As a result, the concept of privacy preserving machine learning (PPML) is proposed, while its form of application varies with different scenarios. Among the research conducted, many studies~\cite{li2020federated,xue2018acies,xue2020sequence} have managed to mitigate the risks of privacy attacks utilizing differential privacy, homomorphic encryption and secure function evaluation, while  these approaches compromise the performance of IoT system, in terms of prediction accuracy, communication cost and the processing speed.
The concept of federated learning was first proposed in 2015~\cite{konevcny2015federated} and may serve as a solution to PPML for IoT without overly compromising the performance. In a generic federated learning process, an initial global model is distributed to the IoT devices, where it is optimised using local samples. The pre-optimized models are then transferred back to server, aggregated to an optimised global model and distributed back to the IoT devices. Since model parameters are transferred, no raw user data is directly exposed to the server. This framework also allows the end devices to be responsible for part of the computation process and could preserve efficiency. However, whether federated learning can obtain an optimal balance between performance and privacy for IoT applications remains unknown. With exposure of models, attacks such as model inversion
and membership inference attack
may still pose threat to user privacy. In addition, while recent studies on federated learning involves communication efficiency~\cite{konevcny2016federated} and performance in accuracy~\cite{hard2018federated} has focused on theoretical analysis, its performance under IoT application context has yet to be practically studied.

In this work, a federated-learning-based system is built to imitate an IoT application scenario. PySyft, a readily available federated learning framework, is employed. Dedicated analysis is conducted on data communication involved during the training process. Performance in terms of training speed, communication cost and plausible privacy concerns of the architecture is evaluated.


\section{Methodology and System Implementation}
\label{sec:preliminarystudy}
\vspace{-0.05in}

The system is developed based on a federated learning application~\cite{flrnn}, which predicts the user's native language based on the given user surname.  A set of text files are supplied as training data, with each file containing the surnames from a certain language. The total size is 10.28MB with 20,074 surnames included. The text is manipulated to tensor format and randomly distributed to the workers. For each epoch, data entry is supplied to the targeted worker and the model feedback from the worker after training is aggregated at the server. A Recurrent Neural Network (RNN) is deployed on each remote worker as initial model for training. Prediction and classification are made on single test text inputs after training accomplishes.

\begin{figure}
	\centering
		\includegraphics[width=3.5in]{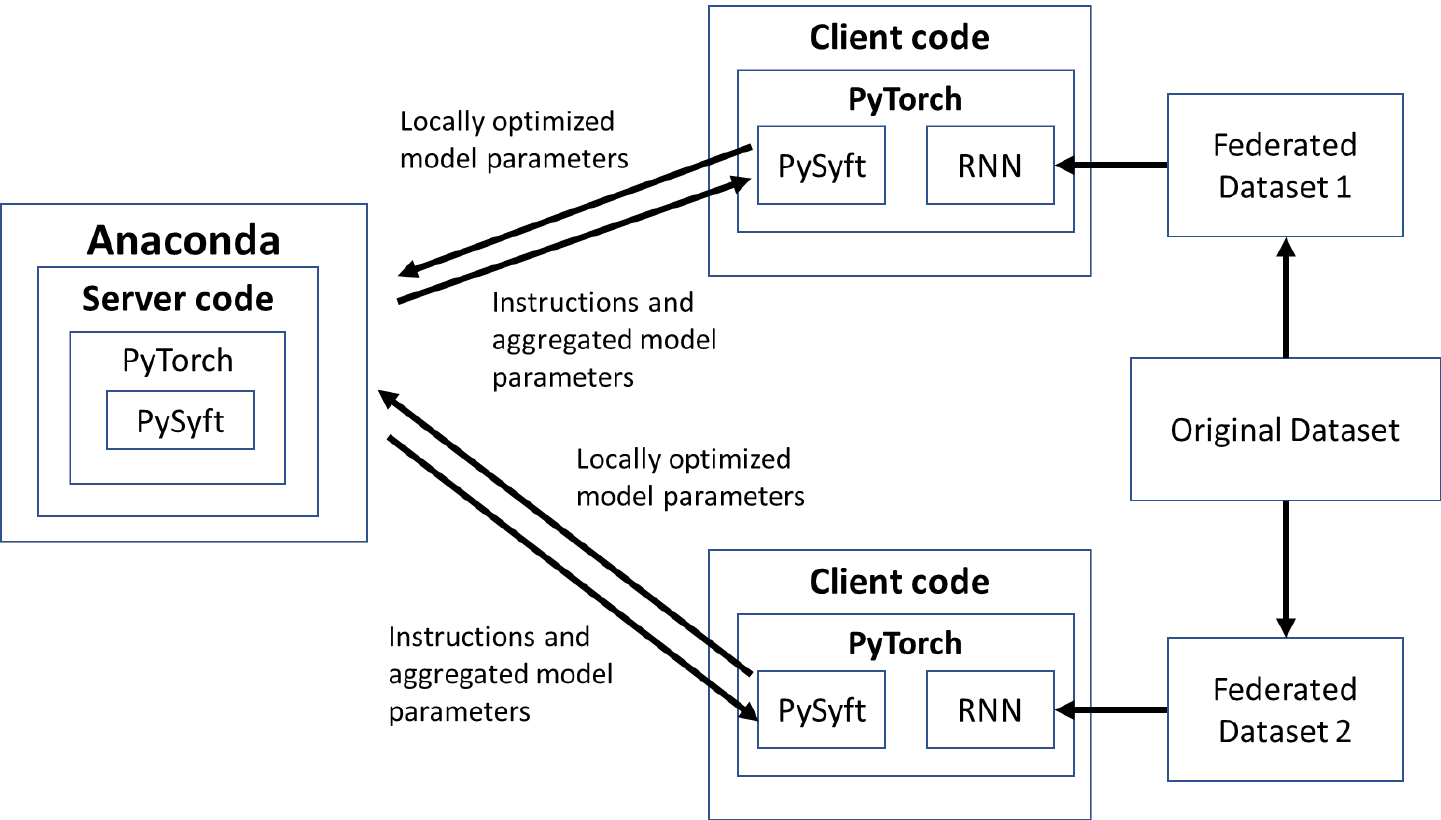}
		\vspace{-0.15in}
	\caption{System Architecture.}
	\vspace{-0.1in}
	\label{fig:Systemmodel}
\end{figure}

The federated learning testbed is set up to closely imitate a practical IoT network (see Figure~\ref{fig:Systemmodel}). It includes two remote workers and one central server. The server and client can be reinforced to communicate across the network. This reserves the accessibility for research and measurement. 
Some key choices and justifications are explained below:

\begin{enumerate}
    \item The project utilises Anaconda on the centralized server to manage python packages. Packages from PyPI can also be installed in conda environment. Strict and explicit version control is required to resolve compatibility issues, and hence multiple conda virtual environments are created to avoid version conflicts.
    \item PyTorch is selected to be the learning framework, which is capable of lower level operations to allow more flexibility than packed high-level frameworks,  and hence is prevalently used for research purpose. The complexity is suitable for a prototype system in this project. PyTorch also supports distributed learning with libraries embedded.
    \item PySyft library allows federated learning to be performed based on PyTorch operations. It enables remote computation process on IoT devices by giving instructions from the centralized server without the knowledge of raw user data. Instructions of PySyft are integrated with a set of python utilities, which ensures the simplicity of implementation. Communications between end devices and server are realized by application layer protocol websocket.
    \item A semantic dataset is tested to make predictions on the language that the given surnames belong to, which aligns the application of federated learning in language recognition (such as Google Keyboard) and allows relatively simple evaluations.
    
\end{enumerate}
	\vspace{-0.2in}


Wireshark is used to analysis communication rate and privacy. Amount of data communication can be intuitively viewed from the ACK and SEQ values of captured TCP packets. This is verified later during the test that the payload is much larger than the headers of the packets, and thus the packet size manifested effectively approximates the amount of data transmitted. 
Privacy of the communication scheme used in this federated learning architecture is analysed by comparing the content of packets received by the receiver code and that captured by Wireshark. While other advanced traffic analysis tools are available, Wireshark is capable of decoding packet to a similar level to that the routers and proxies in the network infrastructure do. PyCharm is used which enables to debug the code and visualize the intermittent variables. The code is executed step by step in debug mode in PyCharm and communication is monitored simultaneously on Wireshark. This ensures that any traffic observed on Wireshark can be mapped to data received at the receiver code in PyCharm. 

\section{Results on Efficiency Test}
\label{sec:protocol}
\vspace{-0.05in}
Test results in terms of training speed, communication cost and data privacy are presented and discussed in this section. But first, the accuracy of prediction  needs to be ensured. Only a few occasional deviations occurred, while most tests have $100 \% $ accuracy. 
Since data distribution is random, one model may become more expertized in predicting one type of text input than the other does. The models from the aggregated model have the same weights, and hence the value from one model can overwhelm that from the other. However, since accuracy remains high, it can be concluded that the fundamental requirement is met.

\vspace{-0.05in}
\subsection{Training Speed}

Measured training speed of the model is tabulated in Table 1. With remote workers configured as virtual workers, objects are transferred between functions without forming websocket packets. The average training time is 7 min 19 seconds. As expected, no traffic is detected on the network interfaces by Wireshark. Compare to a centralized machine learning model, it takes more time to train the model. An important reason is that aggregation needs to be performed as an additional step.  To ensure the accuracy, training is made in a conservative manner such that the workers feed back the model after getting trained on every entry. With 10,000 samples are selected to form the test dataset on each worker, aggregation is thus frequently performed and adds to delay. Another possible reason is that it uses Syft commands to overwrite Torch commands with a hook operation, while it still call Torch within its functions. Hook is only necessary when communication needs to be established. This results in a relatively longer call stack and increases the latency as well. 

\begin{table}[]
\vspace{-0.2in}
\caption{Summary of training time.}
\label{table:trainingtime}
\resizebox{0.8\textwidth}{!}{%
\begin{tabular}{l|cccccc}
\hline
\multicolumn{1}{c|}{Index} & 1                           & 2                            & 3                            & 4                           & 5                           & Average                      \\ \hline
VirtualWorker              & 7m 11s                      & 7m 3s                        & 7m 42s                       & 7m 23s                      & 7m 9s                       & 7m 19s                       \\
WebSocketWorker            & \multicolumn{1}{l}{103m 0s} & \multicolumn{1}{l}{100m 37s} & \multicolumn{1}{l}{100m 15s} & \multicolumn{1}{l}{102m 1s} & \multicolumn{1}{l}{101m 2s} & \multicolumn{1}{l}{101m 23s} \\ \hline
\end{tabular}}
\vspace{-0.2in}
\end{table}

Another scenario is tested where the remote workers is mimicked with WebSocketServers. The average total training time is significantly larger than that in the previous test, which reaches 101 mins 23 seconds. Compared to the previous scenario, Wireshark is capable of detecting websocket packets and TCP packets in Adapter for Loopback Traffic interface. After using the specified port number as filter condition, continuous communication is observed between the remote workers and the central server. The congestion window remains at a large value, indicating a good network condition, which is expected to be observed when no other traffic and delay interference the communication on the same machine. It also suggests that delays are not resulted from congestion. However, over 268,000 packets are observed between the two ports, and for each communication, a series of steps need to be taken to encapsulate or decapsulate the websocket packet, which will be discussed in detail in privacy evaluation section. Hence, it can be concluded that the delay is caused by the overly frequent communication. This is further explained with mechanism of the realization of federated learning by PySyft. PySyft server controls the remote worker by obtaining and operating on pointers to objects on the remote worker. In this case, the server sends out an instruction for every step of model training, including instructions for forward pass, loss calculation, backward pass and parameter tuning, maybe for multiple iterations.  Each of these steps corresponds to one instruction sent via websocket to the remote workers. It can be roughly estimated that the time it takes for a single round of communication takes hundreds of milliseconds. This is further multiplied with the number of training rounds, which sums up to a significant amount of instructions and thus significantly compromised the speed. The delay can be even larger if network congestion is considered.
\vspace{-0.05in}
\subsection{Communication Cost}
\begin{figure}
\vspace{-0.2in}
	\centering
		\includegraphics[width=4.5in]{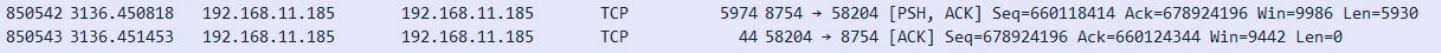}
		\vspace{-0.1in}
	\caption{Total ACK.}
	\vspace{-0.2in}
	\label{fig:totalack}
\end{figure}

\vspace{-0.05in}
Amount of data transmitted over the network can be estimated with the ACK numbers and sequence numbers in packets transmitted both ways. Starting with ACK = 1, SEQ = 1, it is observed that both ACK and SEQ values exceeded 660,000,000 after $20\%$ of training is completed (Figure~\ref{fig:totalack}). Since the amount of data is proportional to the number of training epoches performed, the total amount of data communication for both uplink and downlink to complete the training are 3.3GB respectively for a single remote worker. This result is far larger than the original size of dataset, which is 10.28MB only.

\begin{figure}[!]
  \subfigure[Model delivered to remote worker]{
  \includegraphics[width=4.5in]{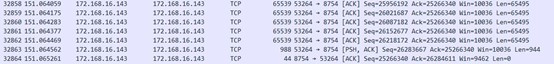}
  \label{fig:txmodeltransfer}}
  
  \vspace{-0.1in}
   \subfigure[Traffic during Link maintenance]{
  \includegraphics[width=4.5in]{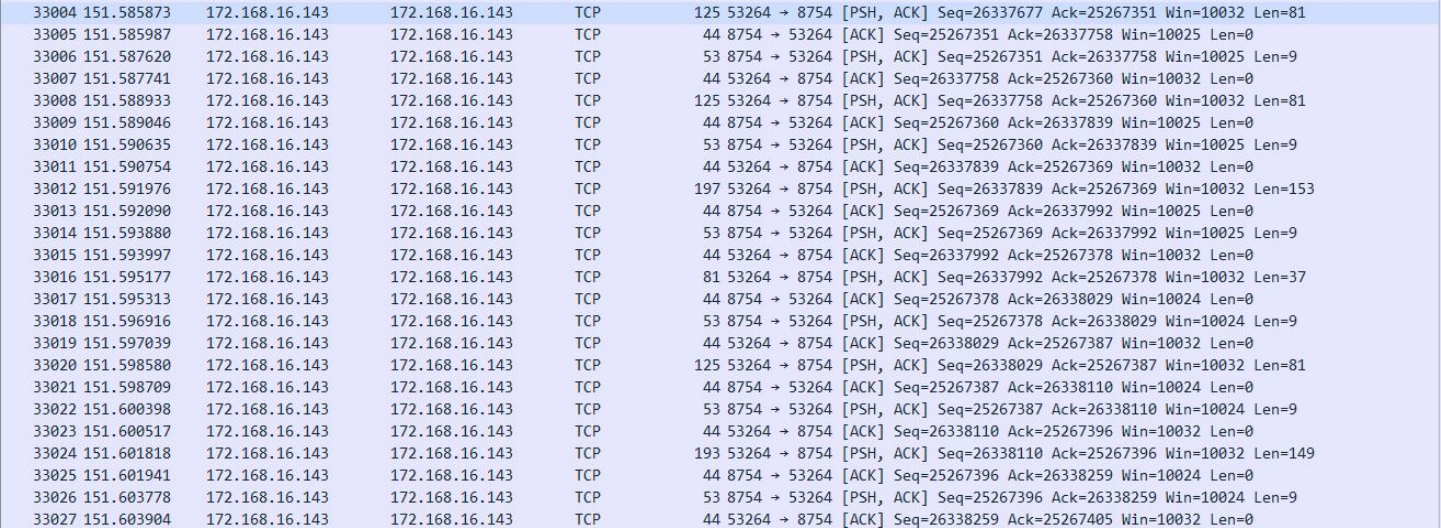}
  \label{fig:training}}
  
  \vspace{-0.1in}
  \subfigure[Model retrieved from remote worker]{
  \includegraphics[width=4.5in]{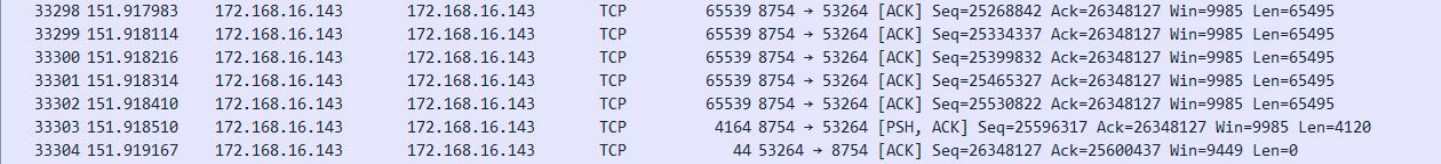}
  \label{fig:rxmodeltransfer}}
  \vspace{-0.15in}
 \caption{Captured Traffic for one Training Epoch}
 \label{capturedtraffic}
 \vspace{-0.1in}
\end{figure}

The observations on pattern of traffic further verified the results obtained above. The size of each model is approximately 330KB, with information about type, structure and parameters of the model included, and is transmitted with 6 consecutive TCP segments (see Figure~\ref{capturedtraffic}). Size of payload is far greater than size of packet header, and therefore total packet size approximates the size of a model. With 10,000 times transmission, the total amount of communication will be 3.3GB (Figure~\ref{fig:totalack}), which agrees with the previous result. The typical length of instructions is between 100 Bytes and 300 Bytes, which does not contribute much to the total data communication. It is further observed that most of the packets have the same length, namely 53B and 44B packets shown in Figure~\ref{capturedtraffic}. They are later identified to be the handshakes of websocket protocol to maintain the connection when waiting for computation to be completed. 

\section{Privacy Analysis}
\label{sec:Privacy Analysis}
\vspace{-0.05in}
 Investigation on privacy of the framework is performed by identifying whether the model  is visible to a third party in the middle of the network, such as a proxy or a router. 

\begin{figure}
\vspace{-0.1in}
	\centering
		\includegraphics[width=3.5in]{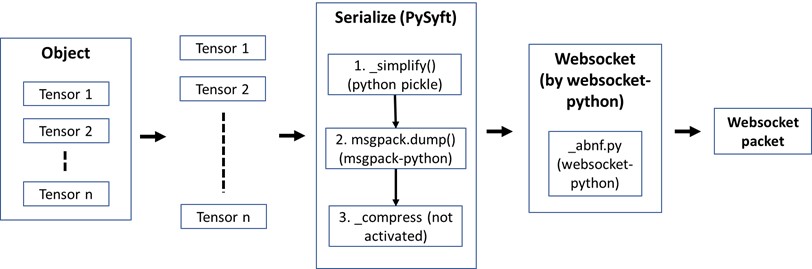}
		\vspace{-0.1in}
	\caption{Process for sending a model.}
\vspace{0.05in}
	\label{fig:txprocessing}
\end{figure}

The process of sending the model is tracked step by step to the lowest level, described in Figure~\ref{fig:txprocessing}. Model parameters are optimized locally by the worker and contained in a python object. The python object is decomposed into tensors and each of them are ‘simplified’ (manipulated to be compatible with the serialization method used in the next step). Then, the tensors are serialized with the default torch serialization strategy, where the python utility pickle is used to convert the object into byte stream. The byte stream is serialized into msgpack format, whereas this conversion process is an implemented python package msgpack available in PyPI
Also, the above process is implemented in function ‘serialization’ in PyTorch library. This byte stream is passed to websocket applications as payload, where packet header specified and added to form the complete packet and then sent to network. The data is then passed to websocket protocol. Optionally, websocket packet can be masked to protect the network infrastructure, so that the user cannot select patterns in the packet.
However, this mask key is added to form a websocket frame and does not give any privacy advantage. Then, after adding the header, the frame is sent onto network without further encryption. Note that the entire process stated above uses methods and formats open to public. This indicates that a websocket packet can be parsed to extract model information. 


\begin{figure}
\vspace{-0.1in}
	\centering
		\includegraphics[width=3.5in]{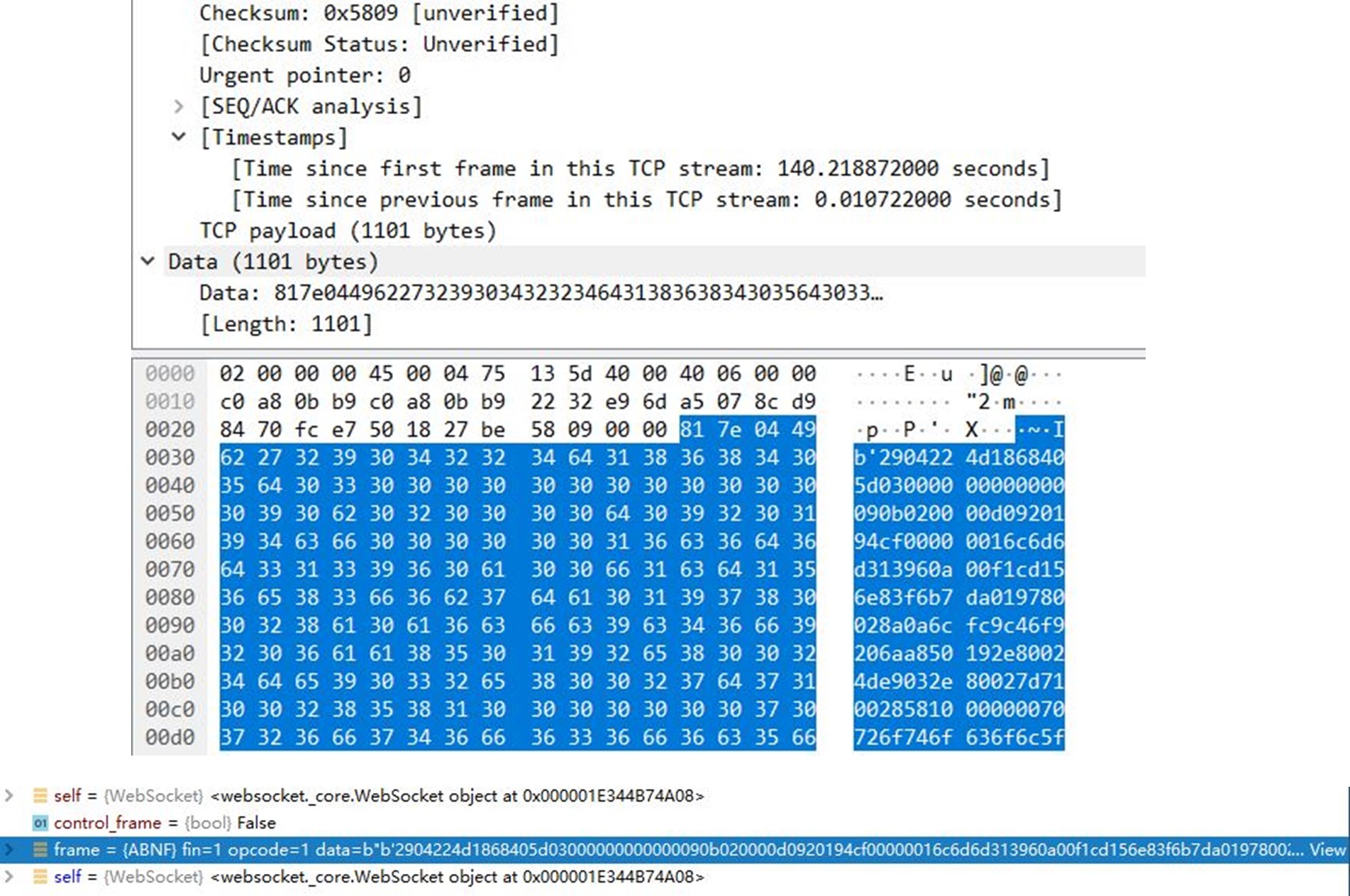}
		\vspace{-0.1in}
	\caption{Payload received by server.}
\vspace{-0.2in}
	\label{fig:serverpayload}
\end{figure}
\begin{figure}
	\centering
		\includegraphics[width=3.5in]{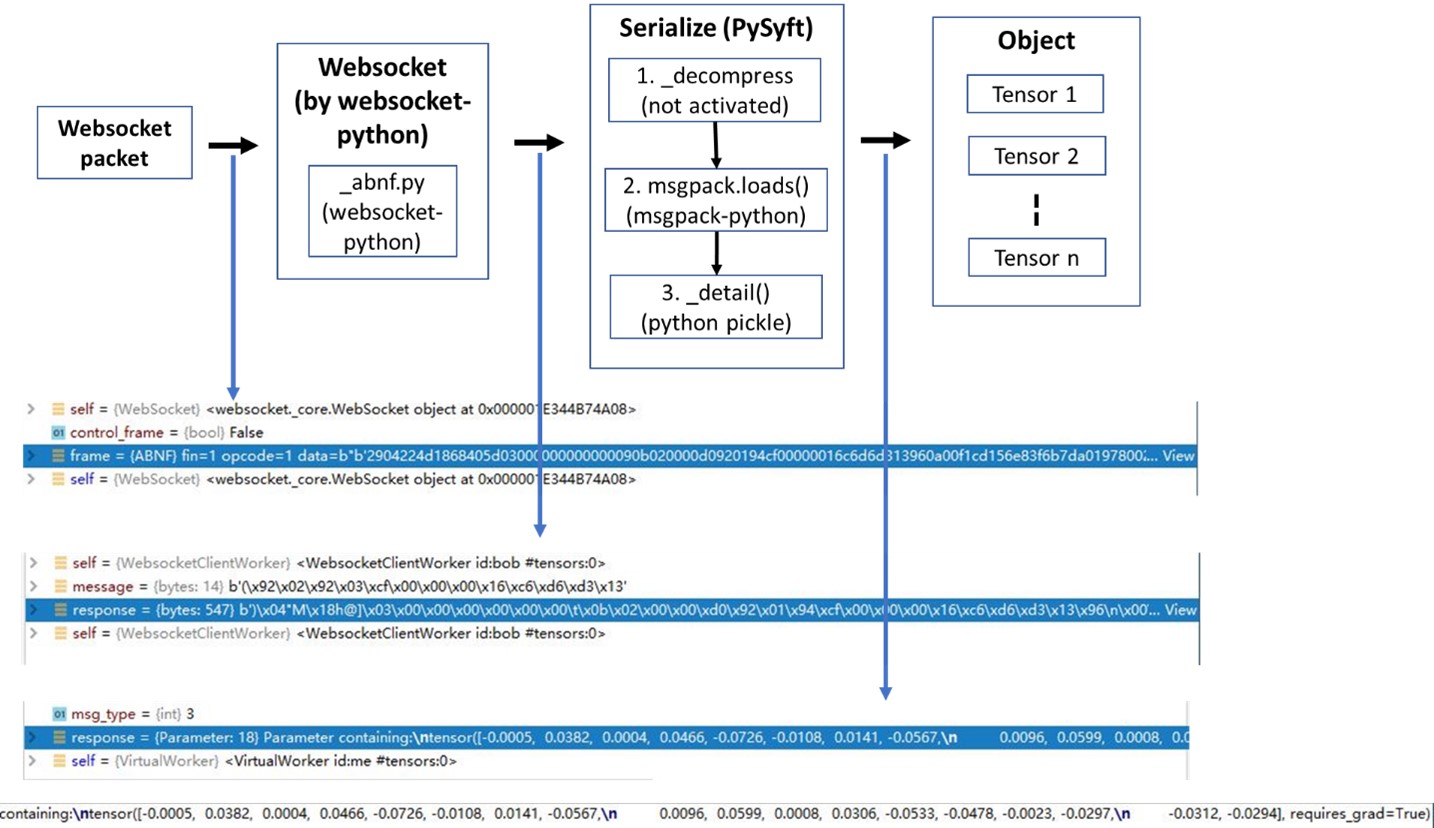}
	\caption{Process for parsing a IP level packet}
\vspace{-0.2in}
	\label{fig:rxprocessing}
\end{figure}

In terms of network level privacy, it can be proved that using the server receiving functions directly can parse all the model information, since it is observed that payload of packets captured by Wireshark and input data to the server receiving function is identical (see Figure~\ref{fig:serverpayload}). 
The process is described in Figure~\ref{fig:rxprocessing}. The attacker can obtain control information in the header of captured packet with the knowledge of websocket frame structure. Then it uses the mask key to unmask all the payload. In example in the figure, no mask is set and therefore the data are identical. 
After that the payload is transferred to hexadecimal format, decompressed and deserialized, and multiple tensors are combined to form the model. These steps all are readily understood, available and implemented by PySyft in ‘deserialize’ function already. That is sufficient to obtain the model structure and parameters transmitted. This implies that all steps are known to parse immediate data received at the server into a model. The process above reveals the vulnerabilities of the communication scheme to man-in-the-middle attack. Any parties in the middle of the network have the same knowledge of transmitted information as the server does, while the server receiving code is completely open, with all process implemented by standard methods and without any encryption. 
There is a safety threat due to exposure of model to other probably malicious parties, which can thus result in loss of privacy.

\section{Discussion}
\label{sec:disc}
Overall, it is observed that the training speed of the federated model is significantly higher than that of traditional centralized machine learning model. The major reason is identified to be the overly frequent communication between the server and remote workers. It is further discovered that communication cost is relatively high. The total amount of data transmitted over the network is much larger than the size of dataset itself. Therefore, it may be concluded that communication between server and remote worker is a paramount problem to be solved before federated learning can be deployed on practical IoT network. More efficient training schemes, for example, to perform aggregation after a group of training iterations may be developed as well, but the prediction accuracy or the quality of the final model achieved remains to be evaluated. Communication frequencies can also be reduced by grouping multiple instructions together, at the cost of flexibility of operations. This technique is already being exploited by PySyft.

In terms of data privacy, it is found that no raw data is transferred through the network. This manifests the advantage of federated learning that raw user data is not directly exposed to a third party. However, the vulnerability of this federated learning architecture is also revealed. Man-in-the-middle attack can directly use receiving functions of PySyft to extract model parameters from packets captured, and possibly deduce user data given models at different stages of training. While the current PySyft only  supports differential privacy over local user data, privacy preserving techniques that can be applied to federated learning methods are yet to be developed. The Homomorphic Encryption decoupled with PySyft may also be enabled for model protection, but it might be faced with even more significant compromise of efficiency performance. To cope with these problems, a possible approach is to embed the concept of the secure multi-party computation, such as shared governance in practice, into a more carefully designed and more efficient instruction and model exchange protocol between the server and workers, which will reduce communication cost and provide more privacy-preserving features.

As for continuing work of this project, practical federated learning frameworks other than PySyft can be evaluated with similar approaches, which may provide different trade-offs between privacy and efficiency performance. Different application scenarios of federated learning can be tested as well, where the testing dataset and type of neural network may be altered with those with higher complexity, coupled with more well-designed training and information exchange strategies, and the subsequent changes in efficiency and accuracy would be an interesting topic to explore. Also, other advanced tools can be used to detect the vulnerabilities of the architecture against other types of attacks and hence formalize more dedicated testing framework. These steps may facilitate better understanding to the mechanism of currently available federated learning architectures and their competence for privacy-preserving machine learning for IoT. The study can be further extended to the test of those federated learning frameworks on microcontrollers for best-effort imitations of practical IoT application scenarios. Resolutions to limited processing resources, power consumption and hardware compatibility at that stage will be a final step to the commercial and industrial realization of PPML enabled IoT. 
\vspace{-0.15in}
\section{Conclusion}
\vspace{-0.05in}
\label{sec:conclusion}
In this project, a federated learning network testbed based on PySyft is deployed and investigated as a candidate solution to privacy preserving machine learning for the IoT. After dedicated evaluations and analysis, it can be concluded that the training speed and communication efficiency of the framework tested is significantly compromised due to the overly frequent transmission of models and instructions across the network. Meanwhile, the system is identified to have vulnerabilities to man-in-the-middle attacks, where user data may be reversely deduced for the model parameters captured at network layer. Even though, PySyft has suggested an innovative approach for multi-party computation, which, with potential enhancement in efficiency and privacy features , could shape the future of privacy preserving machine learning for the IoT.

 \vspace{-0.15in}
 
  \section*{Acknowledgments}
\vspace{-0.05in}

The work has been supported by the Cyber Security Research Centre Limited whose activities are partially funded by the Australian Government’s Cooperative Research Centres Programme.

\vspace{-0.15in}

\bibliographystyle{splncs04}
\bibliography{reference}

%



\end{document}